\documentclass[runningheads]{llncs}
\usepackage{graphicx} 
\usepackage[T1]{fontenc}
\usepackage[colorlinks=true, linkcolor=blue, citecolor=blue, urlcolor=blue]{hyperref}

\usepackage{mathtools}
\usepackage{tabularx}
\usepackage{cleveref}
\usepackage{array}
\usepackage{comment}
\usepackage{enumitem} 
\usepackage{comment,etoolbox}
\usepackage{xargs}
\usepackage{graphicx}
\usepackage[colorinlistoftodos,prependcaption,textsize=footnotesize]{todonotes}
\usepackage[normalem]{ulem}
\usepackage{soul}
\usepackage{xtab}
\usepackage{lipsum} 
\usepackage{changepage} 
\usepackage{marvosym}

\usepackage[misc,geometry]{ifsym}
\usepackage{academicons}

\title{Addressing Privacy Concerns in Joint Communication and Sensing for 6G Networks: Challenges and Prospects} 

\titlerunning{Privacy in JCAS for 6G}

\begin{document}

\author{Prajnamaya Dass\inst{1}\raisebox{0.5ex}{(\Letter)} \and 
Sonika Ujjwal\inst{2}\and
Jiri Novotny\inst{2} \and
Yevhen Zolotavkin\inst{1} \and
Zakaria Laaroussi\inst{2}\and
Stefan Köpsell\inst{1}}

\authorrunning{P. Dass et al.}
%
\institute{Barkhausen Institute, Dresden, Germany\\
\email{\{prajnamaya.dass, yevhen.zolotavkin, stefan.koepsell\}@barkhauseninstitut.org} 
\and
Oy LM Ericsson Ab, Jorvas, Finland\\
\email{ \{sonika.a.ujjwal, jiri.novotny, zakaria.laaroussi\}@ericsson.com}
}

\maketitle

\begin{abstract}
The vision for 6G extends beyond mere communication, incorporating sensing capabilities to facilitate a diverse array of novel applications and services. However, the advent of joint communication and sensing (JCAS) technology introduces concerns regarding the handling of sensitive personally identifiable information (PII) pertaining to individuals and objects, along with external third-party data and disclosure. Consequently, JCAS-based applications are susceptible to privacy breaches, including location tracking, identity disclosure, profiling, and misuse of sensor data, raising significant implications under the European Union’s general data protection regulation (GDPR) as well as other applicable standards. This paper critically examines emergent JCAS architectures and underscores the necessity for network functions to enable privacy-specific features in the 6G systems. We propose an enhanced JCAS architecture with additional network functions and interfaces, facilitating the management of sensing policies, consent information, and transparency guidelines, alongside the integration of sensing-specific functions and storage for sensing processing sessions. Furthermore, we conduct a comprehensive threat analysis for all interfaces, employing security threat model STRIDE and privacy threat model LINDDUN. We also summarise the identified threats using standard common weakness enumeration (CWE). Finally, we suggest the security and privacy controls as the mitigating strategies to counter the identified threats stemming from the JCAS architecture.

\end{abstract}

\section{Introduction} \label{sec:intro}
The evolution of mobile communication, spanning from the inception of 1G to the latest iteration 5G, has reshaped the fabric of human connectivity and interaction. While technologies like beamforming and network slicing have bolstered efficiency, 5G remains primarily communication-focused  \cite{5g_comm_centric}. Looking ahead, 6G aims to surpass mere communication, integrating sensing for a myriad of innovative applications and services \cite{JCAS_IEEE_access}. 
The integration of communication and sensing capabilities, also referred as joint communication and sensing (JCAS), reflects a growing enthusiasm to unlock its capabilities in solving real-life challenges efficiently. In the literature, the term JCAS is alternatively denoted as integrated communication and sensing (ICAS) or integrated sensing and communication (ISAC). Throughout this paper, we use the term JCAS to represent this concept. There are many use cases proposed by 3GPP and other organisations that focus on the potential applications of JCAS across various domains, e.g., autonomous driving, smart city, precision agriculture, industrial IoT, healthcare and telemedicine \cite{hexa_vision,usecase,JCAS_IEEE_access}. Despite its numerous technological benefits and emerging use cases, JCAS introduces various security and privacy challenges \cite{ISAC_Huwaei,JCAS_security_privacy,Hexa_6g_crosslayer_perspective}. Given the incorporation of sensing data, which may contain sensitive personally identifiable information (PII) regarding individuals and objects, JCAS-based applications are increasingly vulnerable to privacy attacks, including location tracking, identity disclosure, profiling, and misuse of sensor data.


In the context of JCAS, the predominant emphasis in current security and privacy solutions lies within physical layer security mechanisms. On the other hand, the privacy mechanisms tailored for independent sensing mechanisms may not be suitable for JCAS scenarios and their particular use cases. The integration of sensing into the current 3GPP architecture necessitates supplementary core network functionalities beyond the sensing management function (SeMF). Additionally, given these added core functions and changes in radio signalling, a distinct threat analysis is imperative, separate from independent communication scenarios. Consequently, the mitigation strategies should include strong security and privacy measures to effectively counter these specific threats posed by JCAS.

Therefore, recognising the sensitivity of JCAS technology and the gaps in existing literature, we present the following contributions in this study.

\begin{itemize}
    \item We conduct a critical assessment of the emergent JCAS architecture, identifying potential security and privacy challenges (Section~\ref{sec:arch-prior-art});
    \item To tackle these challenges, we propose enhancements to the emergent JCAS architecture, introducing new network functions (NFs), interfaces, and data flows (Section~\ref{sec: prop_jcas_arch});
    \item We perform a comprehensive threat analysis of the proposed JCAS architecture, considering the introduced interfaces and components. Our analysis utilises both STRIDE and LINDDUN threat models to cover security and privacy risks comprehensively (Section~\ref{sec:threat_analysis});
    \item Finally, we suggest security and privacy controls to counter the identified threats (Section~\ref{sec:discussion}).
\end{itemize}

\section{Related Works} \label{sec:related_works}

Due to the widespread interest in JCAS, researchers have extensively explored various facets of this paradigm. Architectural concepts, primarily addressing anticipated alterations in the core network functionalities to incorporate sensing alongside communications, have been discussed in works \cite{Arch_Nokia,arch-liu23,Arch_US_patent,ISAC_Huwaei}.

Many existing works focus on the security aspects of JCAS at the physical layer. In \cite{secure_jcas}, the study addresses the challenge of reducing information leakage between communication and sensing functionalities within systems that concurrently perform both operations. In \cite{spoofing_jcas}, the authors conduct a comprehensive security assessment of spoofing attacks in an mmWave radar-based sensing system for autonomous vehicles, incorporating the development and execution of tangible physical layer attack and defence tactics within a cutting-edge mmWave test environment. A spatio-temporal spoofing detection mechanism leveraging MIMO beamforming, was proposed in \cite{spoof_radar} to mitigate spoofing against automotive radars. 

In \cite{poster_jcas_privacy}, the authors emphasised the privacy issues due to sensing activities, such as activity monitoring of sensing targets, eavesdropping attacks from the sensing signals, and false data injection attacks. They suggest the creation of a cross-domain technique for sensing and localisation to accurately recognise human activities, becoming less dependent on location. The work \cite{personal_sensing}, explores privacy concerns surrounding personal sensing, where individuals utilise devices to monitor their activity, location, and environment, and proposing strategies to enhance privacy sensitivity in personal sensing technologies. Participatory sensing allows users to collect and share data through their mobile devices. The work \cite{survey_sensing_privacy}, investigates the privacy concerns due to the use of multi-modal sensors in mobile phones, evaluates existing privacy solutions, and discusses potential countermeasures to safeguard user privacy. In similar works \cite{participatory_sensing,Participatory_sensing_SPEAR}, the authors addresses privacy protection by defining requirements, proposing an efficient infrastructure for mobile users, and discussing open problems and research directions.

\section{Emergent JCAS Architecture} \label{sec:arch-prior-art}

\begin{figure}
    \centering
    \includegraphics[width=1\linewidth]{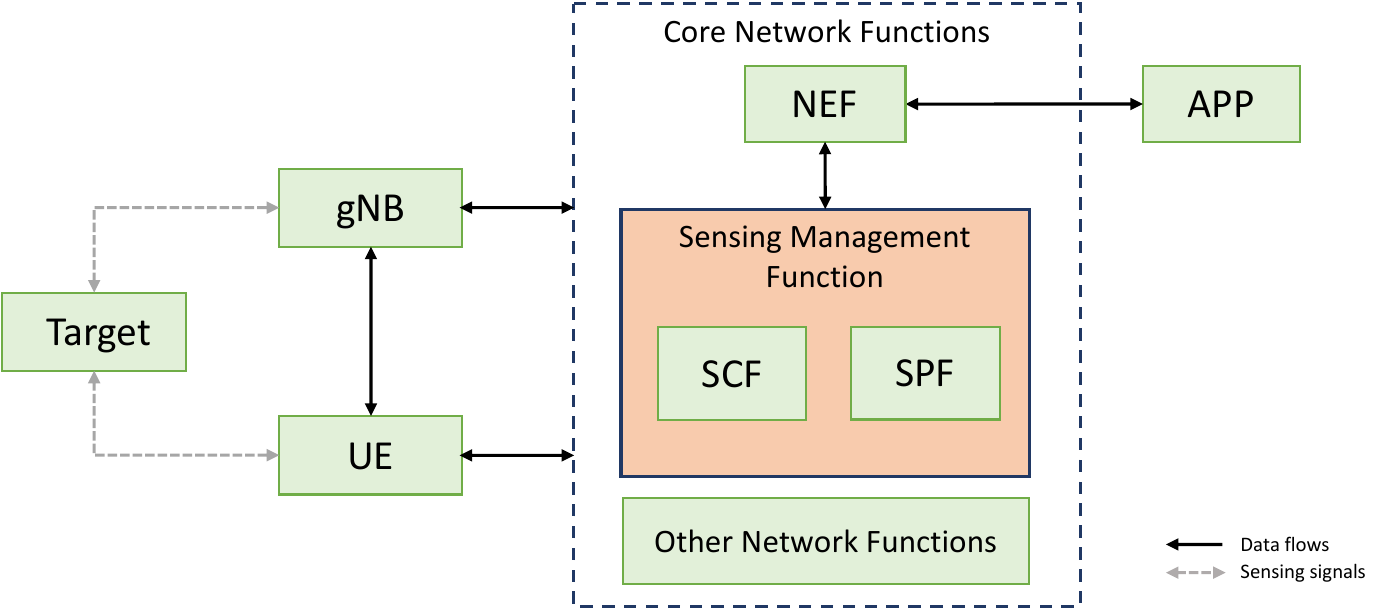}
    \caption{Simplified depiction of the emergent JCAS architecture in a mobile network based on prior art.}
    \label{fig:JCAS_arch}
\end{figure}

\begin{figure}
    \centering
    \includegraphics[width=0.95\linewidth]{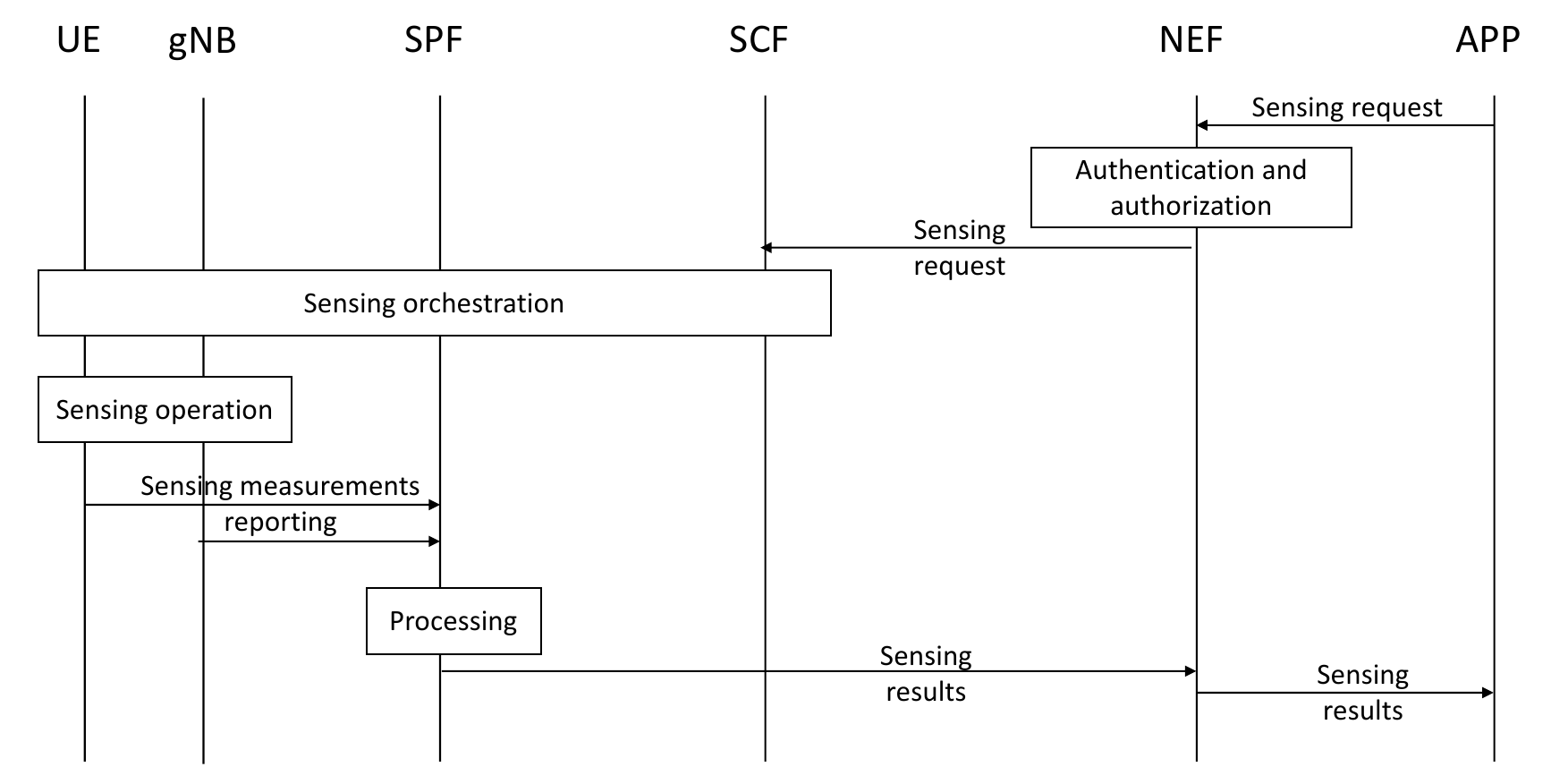}
    \caption{Simplified sequence diagram for JCAS operation based on prior art.}
    \label{fig:jcas-arch-prio-seq}
\end{figure}

Aligning with the architectural concepts outlined in \cite{Arch_Nokia,Arch_US_patent,ISAC_Huwaei,arch-liu23}, we use the emergent JCAS architecture depicted in Fig.~\ref{fig:JCAS_arch} as a baseline for further discussion. To deal with the sensing activities, an additional network function -- sensing management function (SeMF), sometimes referred to as Sensing function (SF) is added to the core network. The SeMF has two main components, namely, sensing control function (SCF) and sensing processing function (SPF). SCF leverages the control plane to receive sensing requests and orchestrate necessary actions with other network functions and entities, such as sensing enabled base stations (gNBs) and UEs. Sensing requests originating in an application (APP) are authenticated and authorised by the network exposure function (NEF). Alternatively sensing requests could be invoked directly from an application function (AF). After sensing signalling between relevant gNBs and UEs the collected sensing measurements are reported to SPF, using for example data plane \cite{arch-liu23}, for processing to gain a semantic understanding of the physical environment depending on the sensing task. SPF may perform tasks such as data aggregation, signal processing, object classification, and anomaly detection to enhance situational awareness and provide results to the application through the NEF. The above described interactions are shown in Fig.~\ref{fig:jcas-arch-prio-seq}.


\subsection{Privacy and Security Challenges in the Emergent JCAS Architecture} \label{subsec:privsec-challenges}


The challenge of a privacy-preserving JCAS architecture remains unresolved, as the nature of sensing data differs from communication data, posing difficulties in integrating sensing into existing 5G/5G-Advanced frameworks. Unlike communication data, sensing data, akin to other sensor data, primarily captures observations about the physical environment and its objects, which may not always be directly linked to a subscriber. For instance, in JCAS-assisted automotive manoeuvring and navigation system, a vehicle user may be tracked by other vehicles and the service provider. The service provider may control the sensing unit of the vehicle and use other vehicle features without user's consent. On the other hand, a vehicle with sensing capabilities can also sense the targets in sensitive or restricted zones. Considering that sensing data may potentially include personal data, whether directly or indirectly, the integration of JCAS in 6G should adhere to the principles outlined in the GDPR on how to collect, use, transfer, store, and dispose the sensing data. 


Introducing the additional sensing functions into the core network for JCAS poses several threats. Firstly, it expands the attack surface of the network, providing more avenues for malicious actors to exploit vulnerabilities. Secondly, the increased volume and diversity of sensing data collected through these sensing functions heighten the risk of data breaches and privacy violations. 
Additionally, attackers could exploit vulnerabilities in one sensing function to breach trust boundaries and gain unauthorised access to other network components. Inadequate enforcement of trust boundaries may lead to data leakage between sensing functions or network domains, compromising the confidentiality and integrity of sensitive information. Robust access controls, authentication mechanisms, and continuous monitoring are essential to mitigate threats associated with trust boundary violations and ensure the security of the JCAS system. Addressing these threats requires robust security measures, stringent privacy protections, and effective management practices to ensure the resilience and integrity of the JCAS system. 



\section{Proposed Architectural Enhancement} \label{sec: prop_jcas_arch}

In this section we describe the architecture that is evolving from the prior art architecture from section~\ref{sec:arch-prior-art} and introducing new concepts in attempts to address some of the privacy challenges covered in section~\ref{subsec:privsec-challenges}.

\begin{figure}
    \centering
    \includegraphics[width=0.8\linewidth]{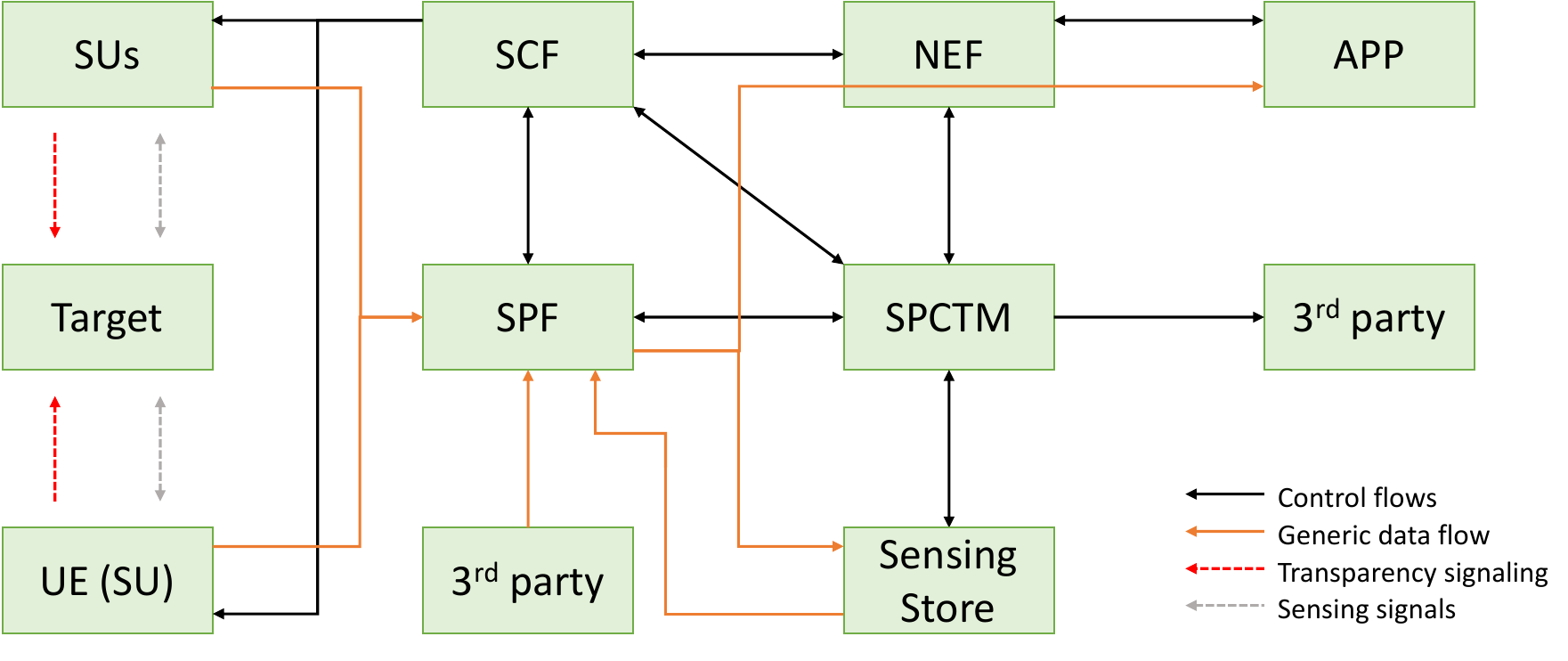}
    \caption{Bird's-eye view of the assumed data flow diagram for this work.}
    \label{fig:dfd-bird}
\end{figure}

Fig.~\ref{fig:dfd-bird} shows a bird's-eye view of the proposed architecture. Similarly to prior art, the NEF brings network capabilities to applications and we retain the SCF and the SPF. However, we propose a complementing component called the sensing policy, consent, and transparency management (SPCTM), which governs sensing privacy, as well as a sensing store to hold persistent data, policies, and logs tailored to the requirements of the sensing function. We propose that the sensing store is an NF-specific solution at this stage, which is not uncommon in current 5G deployments \cite{etsi-5gs}, in order to grasp requirements and possible threats against such an approach. Future studies may investigate the possibilities of reusing existing data store solutions in 5G architecture such as UDM, UDR, UDSF etc. We use the term sensing unit (SU) to refer to the sensing radio component that can be independent (e.g., at the gNB) or part of the user equipment (UE) \cite{usecase}. Note that in the later case, additional UE-based controls govern access to the sensing data to put the user under control. SUs transmit sensing signals to the targets and then capture the reflected signals from them, additionally SUs may be instructed to notify sensing targets about the current sensing session over an air interface such as broadcast. Captured sensing signals are interpreted into sensing measurements which are then sent to the SPF for further processing. The processed measurements, named sensing results, are then disclosed via the NEF back to the application. The interfaces needed towards communication system, depending on the level of integration \cite{hexa-d33}, are not part of the scope of this document. The working assumptions and detailed descriptions of each component are provided in the following subsections.

\subsection{The Network Exposure and Application}\label{sec:NEF-APP}

\begin{figure}
    \centering
    \includegraphics[width=\linewidth]{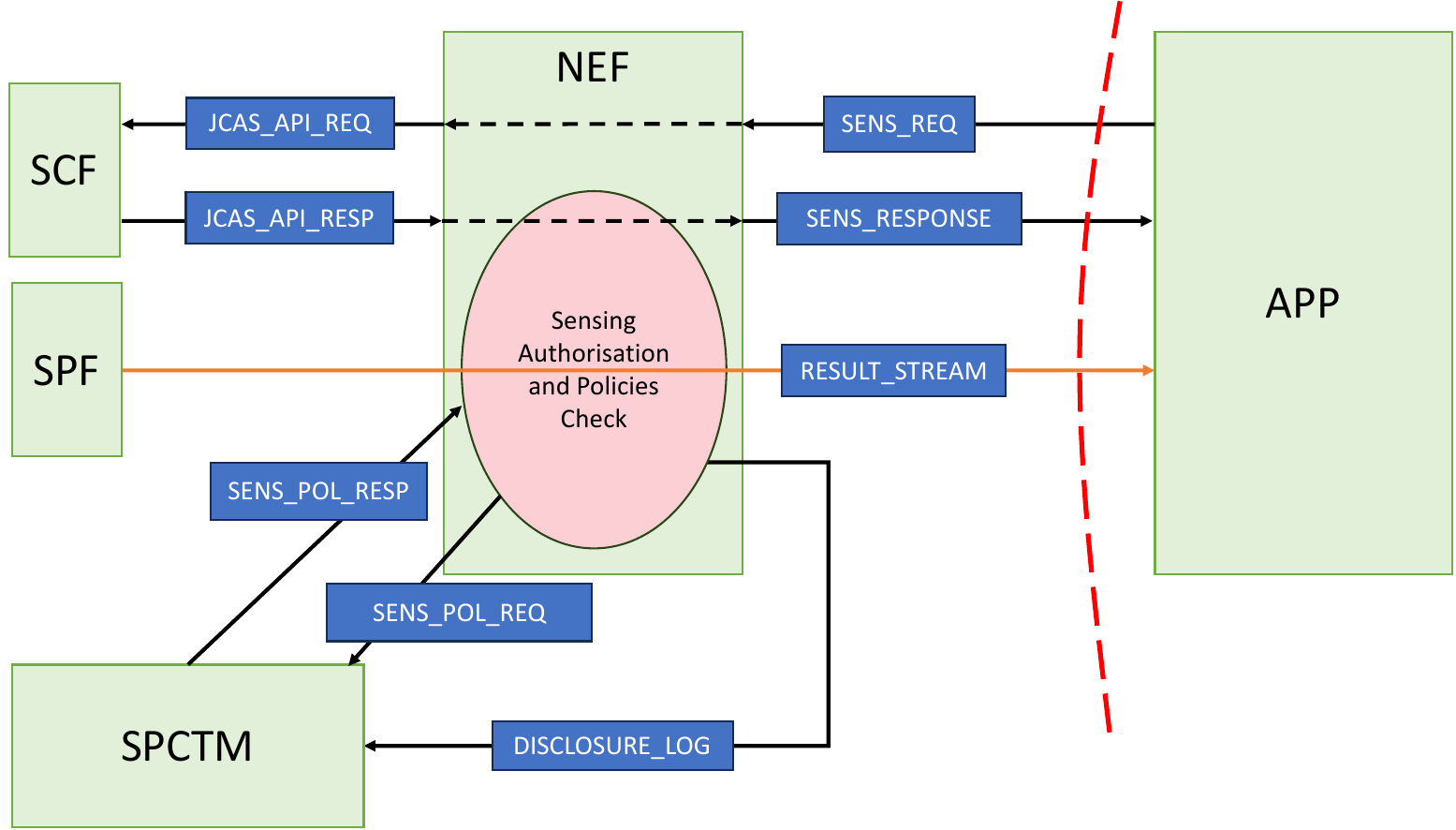}
    \caption{Proposed control and data flows from and to the NEF.}
    \label{fig:APP_NEF}
\end{figure}


The NEF is an essential part of 5G architecture, enabling secure third-party access to 3GPP network services and capabilities, while enforcing policies on data sharing through APIs \cite{3GPP_TS_23.502}. Fig.~\ref{fig:APP_NEF} shows how the NEF enables sensing services, including authentication and policy-based authorisation of sensing applications, with help from the sensing authorisation and policies check component.


Sensing is initiated with the sensing request (\texttt{SENS\_REQ}), which at the very least contains descriptions of the sensing target and sensing results, which differ depending on the given use case. For example, for an early collision warning application on a highway \cite{usecase}, this could be the geo-location as the target and an event-like notification as the response. However, depending on the scenario, the request is expected to be elaborate, containing other fields such as quality of service (QoS), quality of data (QoD), periodicity, and more. In return, the application receives a sensing response (\texttt{SENS\_RESPONSE}) indicating the status — success or failure. The response may carry a sensing result or provide information on how to obtain the results, for example, a web socket to listen in for stream-like sensing results or events (\texttt{RESULT\_STREAM}).

The NEF requests necessary policies (\texttt{SENS\_POL\_REQ}) from the SPCTM, receiving sensing specific authentication and authorisation details (\texttt{SENS\_POL\_RESP}) for the sensing application, such as geo-location permissions and result granularity. The SPCTM assigns a reference (e.g., policy ID) for policy tracking, shared across the NEF, SCF, and SPF. These data flows occur initially and may repeat periodically to ensure proper disclosure of sensing responses and results to the application.



The NEF records data disclosures for transparency, complying with legal standards like GDPR via the \texttt{DISCLOSURE\_LOG} flow. It logs for example recipient identities, data descriptions, disclosure purposes, obligations, timestamps, and applied policies.

After the sensing request has been authorised and the initial set of policies has been established, the NEF proceeds to relay the request to the SCF using a lower-level style API (\texttt{JCAS\_API\_REQ}). In response (\texttt{JCAS\_API\_RESP}), the NEF receives a detailed message from the SCF indicating whether the request was successfully processed or if it encountered a failure. 

\subsection{Sensing Policy, Consent, and Transparency Management (SPCTM)}\label{sec:spctm}

The SPCTM function is designed to administer privacy controls and extend support to other NFs participating in the sensing ecosystem, enabling them to adhere to privacy preservation principles. The SPCTM framework proposed herein does not encompass the privacy considerations for all conceivable use cases; instead, it proposes a foundational model.

\begin{figure}[!ht]
    \centering
    \includegraphics[width=\linewidth]{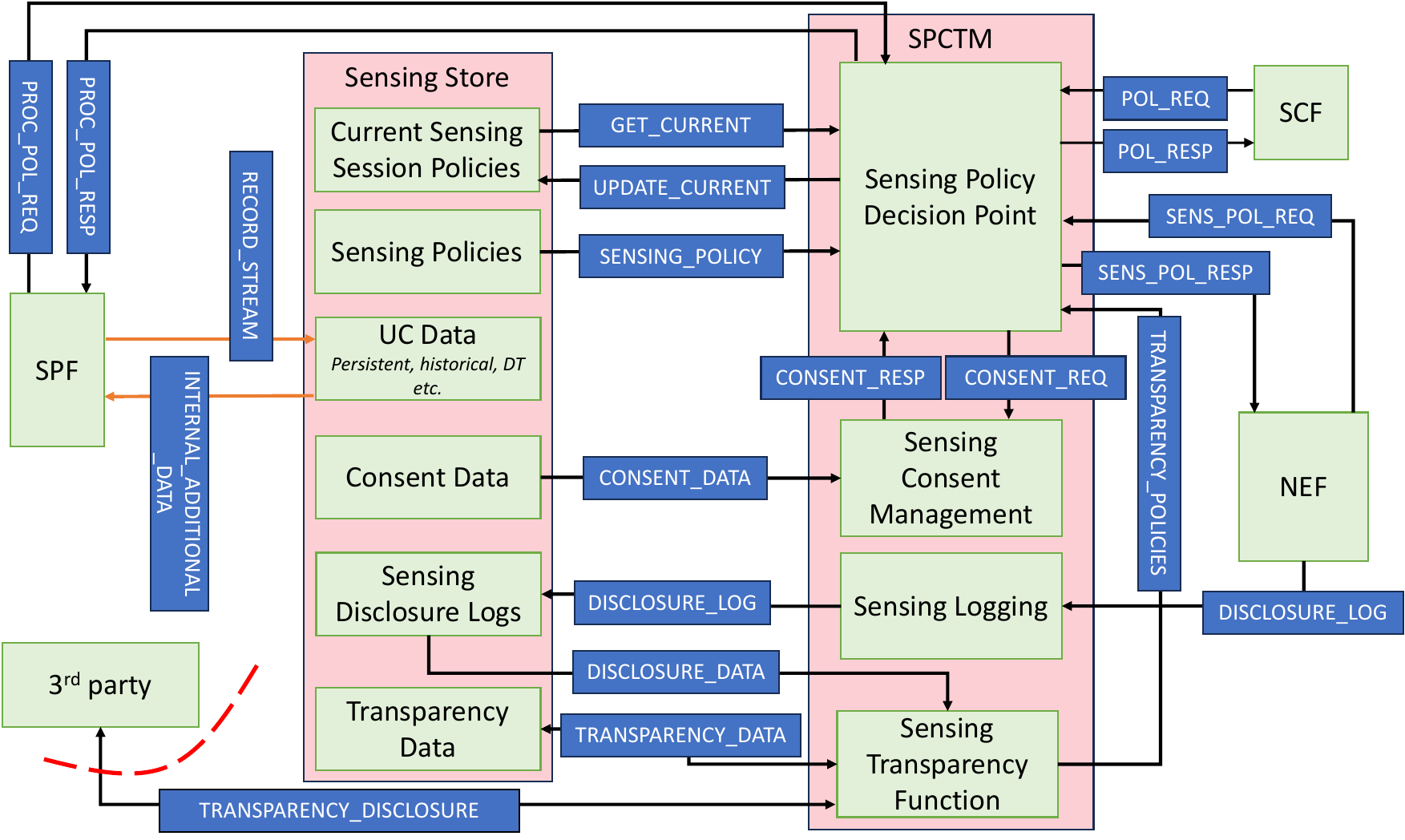} 
    \caption{SPCTM, governing sensing privacy, and Sensing Store interactions with each other and the rest of the proposed system.}
    \label{fig:spctm}
\end{figure}

The components and interfaces of the SPCTM are illustrated in Fig.~\ref{fig:spctm}. The sensing policy decision (SPD) point functions as a central hub for gathering and consolidating sensing policies, consent information, and transparency guidelines, and subsequently disseminating this aggregated information to other NFs. It also maintains a record of the current policies applicable to active sensing sessions. It is presupposed that these policies and associated consent or transparency data, although predetermined, may be subject to change over the course of a sensing session. The SPD point is tasked with the timely notification and updating of relevant components and NFs to reflect these changes, acting as the primary interface between the SPCTM and the remainder of the system. Furthermore, the SPD point is charged with negotiating current privacy policies with the various NFs involved in the sensing process.

The sensing consent management component bears the responsibility for handling consent data and supplying it to the SPD point. Recognising the necessity of such a component is vital for supporting a broader array of sensing applications. However, the precise technical methods for obtaining and managing the consent of all stakeholders are topics for further studies.

The Sensing Logging function interfaces with the NEF, as previously outlined in Section~\ref{sec:NEF-APP}, to facilitate disclosure logging that adheres to transparency requirements. These requirements are provided by the sensing transparency function, which is also responsible for disseminating information on how sensing sessions should be communicated to the affected sensing targets. Options under consideration include directing the SCF, which then appoints SUs, to emit a transparency notification, potentially through a mobile network broadcast, or alternatively, recording the identities of the sensed targets and providing notifications post-sensing procedure completion.

The \texttt{TRANSPARENCY\_DISCLOSURE} interface, between sensing transparency Function and a third party, is designed to provide essential transparency to the sensing targets. It allows them to observe what type of information has been disclosed, to whom, and for what purpose. Additionally, this interface could be managed by a trusted third party or used to support potential audits.

\subsubsection{Sensing Store}

is a storage specific to the sensing function which the proposed architecture in this document employs. This store holds all required persistent data, which we categorise into the following types for the purpose of this document.

\begin{enumerate}
    \item Sensing Policies: Policies governing permissible sensing types, geographic restrictions, disclosure requirements, granularity standards, and privacy are consolidated by the SPD point. These policies provide essential guidelines for authorisation and disclosure to the NEF, control for the SCF, and data handling for the SPF.
    \item Consent Data: This includes consents from sensing targets, where consent is the legal basis for data collection. It covers user permissions for sensing activities on their devices and the extent of their participation. Managing these consents is complex due to the indirect, potentially sensitive, and large-scale nature of sensing data.
    \item Current Sensing Session Policies: These reflect the latest aggregated policies for authorisation, disclosure, control, and processing in active sessions. The storage facilitates access to relevant policies, consent data, and session information, along with any policies composed by the SPD point.
    \item Use Case (UC) Data: Persistent data storage for a specific use case, such as environment maps or historical records, should exclude PIIs and contain only sanitised data. Sensitive data management is addressed separately within the SPF (Section~\ref{sec:spf-sessions}).
    \item Sensing Disclosure Logs: Records of data disclosure detail recipient identities, data descriptions, disclosure purposes, obligations, timestamps, and the policies enforced during the process.
    \item Transparency Data: Policies outline how the collection and processing of sensing data are communicated to impacted individuals and how disclosure is logged and shared with relevant parties, aligning with the mentioned \texttt{TRANSPARENCY\_DISCLOSURE} interface.
\end{enumerate}

\subsection{The Sensing Control and Orchestration}

\begin{figure}
    \centering
    \includegraphics[width=\linewidth]{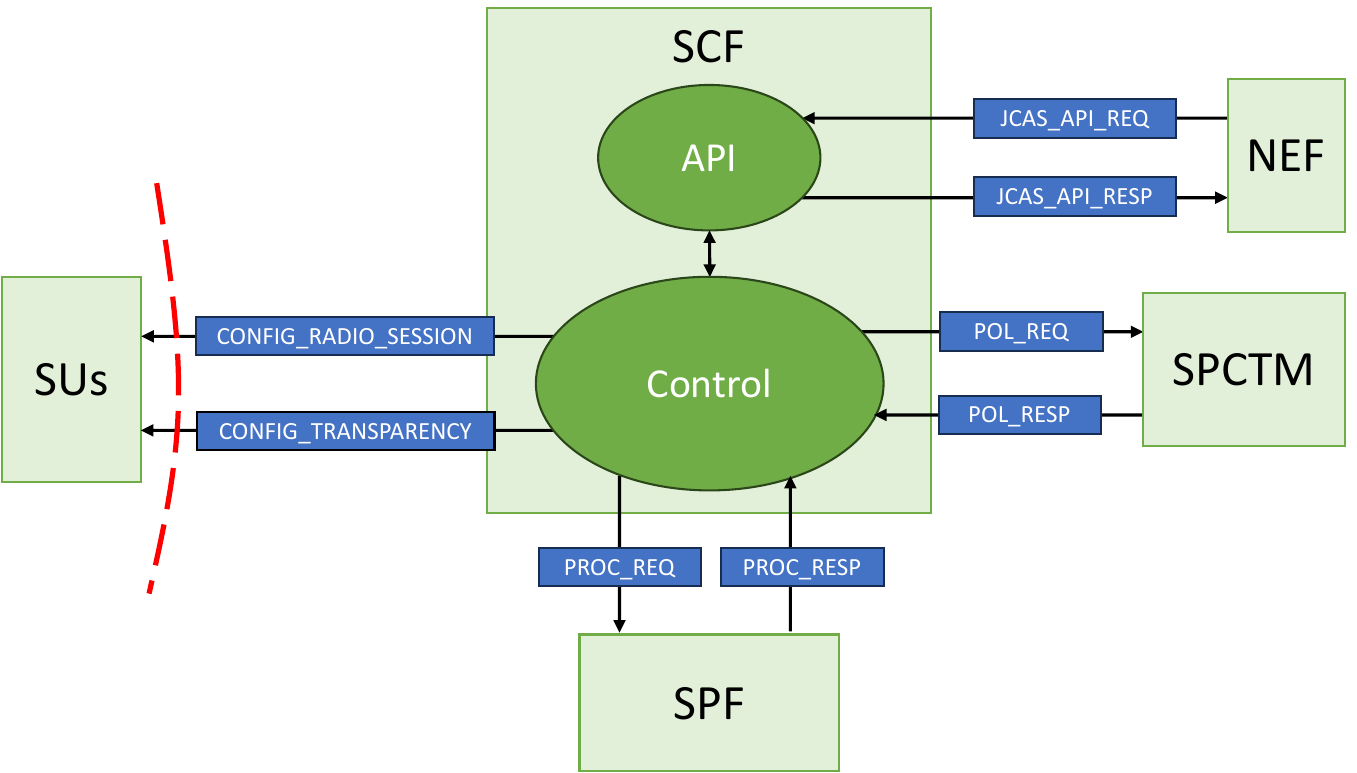}
    \caption{Control flows associated with SCF for managing and orchestrating other NFs to provide requested sensing results.}
    \label{fig:scf}
\end{figure}

Similar to \cite{Arch_Nokia,arch-liu23}, the SCF operates within the control plane of the JCAS framework, as illustrated in Fig.~\ref{fig:scf}. The internal API of the SCF facilitates the flow of sensing requests (\texttt{JCAS\_API\_REQ}). The SCF aggregates multiple requests that can be fulfilled within a single procedure or session, or segregates a single request that cannot be fulfilled within one session, into the necessary set of sensing tasks for measurements and processing.

Upon successful negotiation with the SPCTM over the control policies (via \texttt{POL\_REQ} and \texttt{POL\_RESP}), the SCF requests processing resources for the given task from the SPF using a processing request (\texttt{PROC\_REQ}). The \texttt{PROC\_REQ} includes an estimate of the type, size, and frequency of incoming data, priority or criticality levels, and other necessary parts for processing related to the results required (type, periodicity, reporting style – single, stream, or event) and references to current policies. The \texttt{PROC\_RESP} is a processing response that includes a success or failure indicator, ingress points definitions (e.g., IP address), and optionally, egress points definition (IP address, socket). SCF then sends control parameters to the relevant SUs using the \texttt{CONFIG\_RADIO\_SESSION} command. The SUs execute the sensing operation and send the raw or optionally pre-processed sensing data back to the SPF. The SCF also manages the trade-off between communication and sensing services by efficiently allocating resources. It can potentially fail a request if sensing measurements cannot be obtained as desired, or if the SPF fails to secure resources for needed processing.

The introduction of SPCTM supports more heterogeneous aggregation of sensing tasks by applying correct policies and resolving potentially conflicting ones. The \texttt{POL\_RESP} data contains the current set of control-related policies such as granularity recommendations (time and space), transparency signal information and more. The communication between the SPF and SPCTM is envisioned as a negotiation sequence, allowing the SCF and the SPD point to agree on a solution. The SCF needs to bundle requests together, and the SPD point needs to support this by bundling and resolving relevant policies. Furthermore, the SPD point may need to reflect changes in control policies to processing policies. For example, if a higher granularity is used for sensing measurements, the processing pipeline should compensate as soon as technologically possible. The received indications regarding transparency signalling in \texttt{POL\_RESP} towards affected sensing targets are instructed in \texttt{CONFIG\_TRANSPARENCY} to SUs.

\subsection{Sensing Units}

A sensing unit (SU) is a key component in the JCAS system. It is a radio unit or radio node that possesses the capability to perform a variety of functions such as transmitting and receiving radio signals specifically for sensing purposes, processing of these radio signals, and conducting sensing measurements, among others. The SU may be equipped with or connected to one or more internal or external antennas. In some cases, it may share antennas with other nodes, for example, with gNB or UE. The specific technical solutions and hardware employed by the SU, while important, are not the focus here. Instead, the primary consideration is the SU's ability to be instructed to transmit or receive specific sensing signals at specific times.

\begin{figure}
    \centering
    \includegraphics[width=0.8\linewidth]{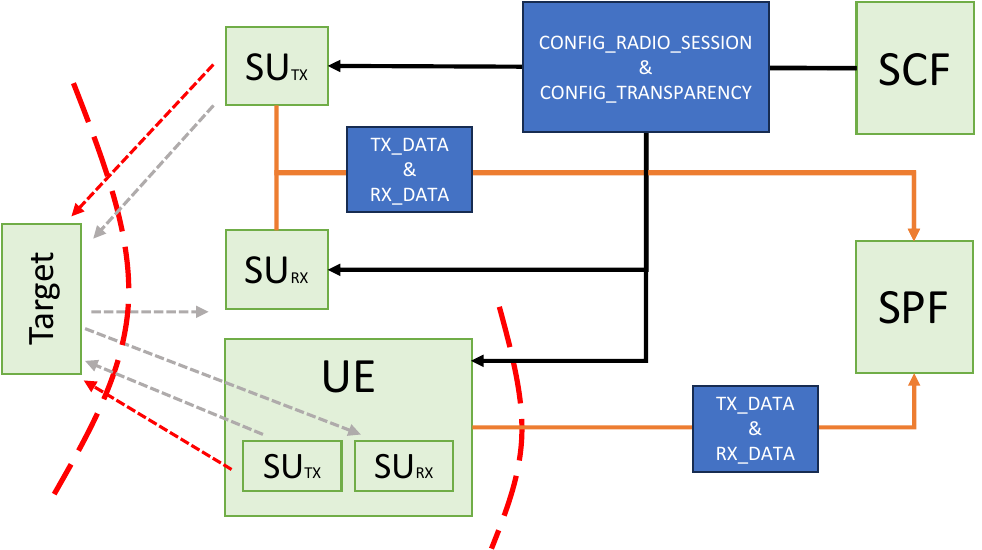}
    \caption{Control flows (\texttt{CONFIG\_RADIO\_SESSION} and \texttt{CONFIG\_TRANSPARENCY}) from the SCF to SUs and data flows (\texttt{TX\_DATA} and \texttt{RX\_DATA}) from SUs to the SPF.}
    \label{fig:sus}
\end{figure}

As illustrated in Fig.~\ref{fig:sus}, the scope is extended to UE-based SUs. Note that these SUs are not directly controlled by the network. Instead, the UE will implement similar privacy controls as the ones in the network (especially SPCTM) (see Section~\ref{secuesu} for more details). Therefore only indirect control is possible, which is governed by user decisions and policies. The controls are omitted in Fig.~\ref{fig:sus} for readability. They are presented in Fig.~\ref{fig:su-control}.

\subsection{Sensing Processing Function}

\begin{figure}
    \centering
    \includegraphics[width=\linewidth]{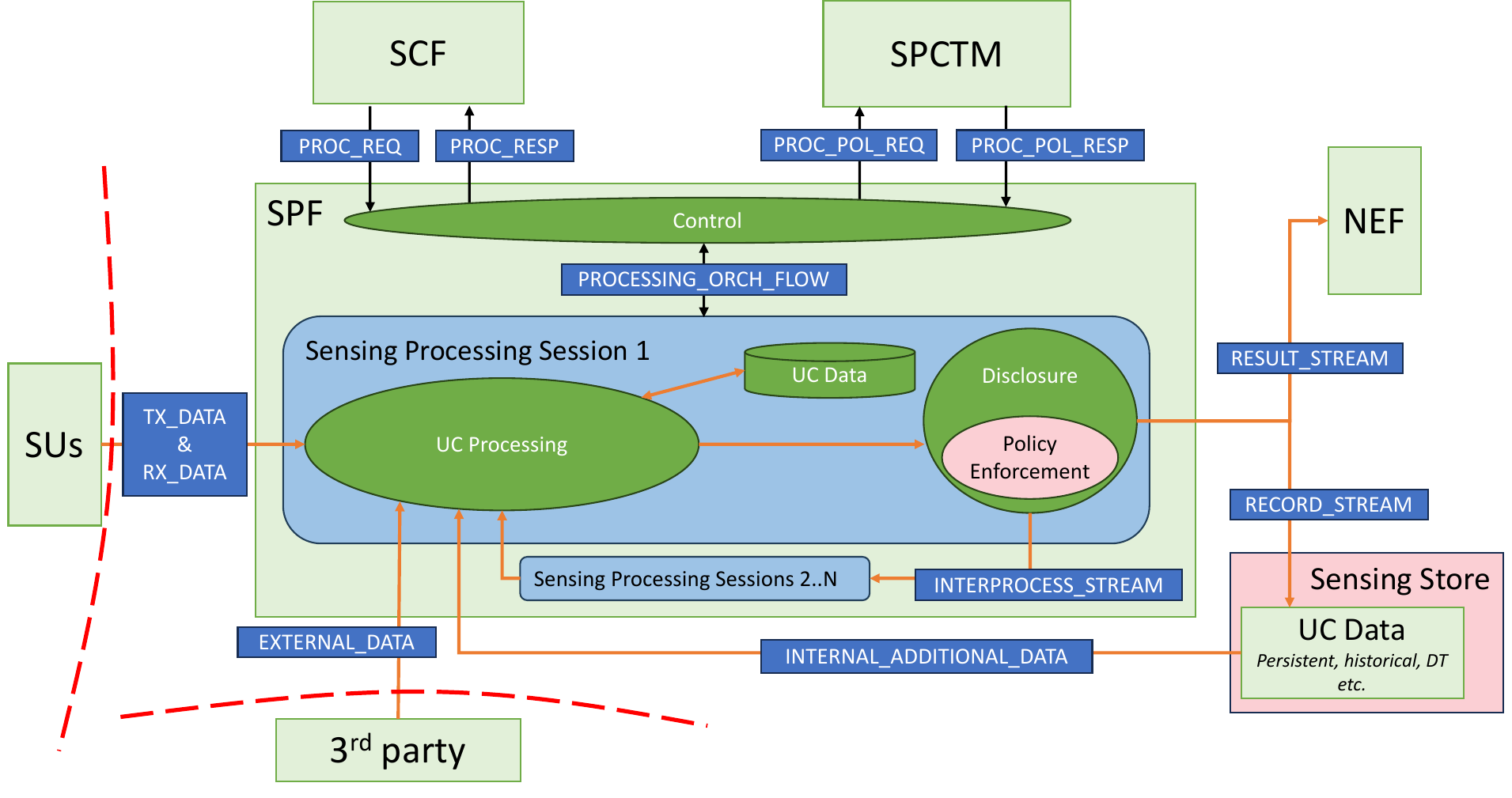}
    \caption{Control flows between SPF, SCF, and SPCTM for sensing data processing.}
    \label{fig:spf}
\end{figure}

The SPF handles raw and pre-processed sensing data from SUs, converting it into usable results for applications delivered via the NEF. As shown in Fig.~\ref{fig:spf}, SPF includes a control component that responds to processing requests (\texttt{PROC\_REQ}) from the SCF, which provides necessary details for processing the data. This component orchestrates processing sessions, managing the required resources for computation and storage to meet the request.

The \texttt{PROCESSING\_ORCH\_FLOW} carries all the orchestration information for SPF sessions, ensuring the processing aligns with the sensing request or task. Once the SPF session is established, it sets up the ingress point for incoming data. The SPF control then updates the SCF with the fulfilment status in the \texttt{PROC\_RES} message. If successful, it also relays ingress (e.g., IP address and port) and potentially egress point details for accessing results.

Like the SCF, the SPF communicates with the SPCTM to receive and enforce the latest processing policies. This interaction occurs initially upon receiving the processing request and subsequently whenever applicable policies change, requiring adjustments in active processing sessions. The \texttt{PROCESSING\_ORCH\_FLOW} may include directives to initiate, end, or modify a session in response to changes in sensing or processing policies.

\subsubsection{Sensing Processing Sessions}\label{sec:spf-sessions}
are transient instances providing necessary processing for fulfilling sensing requests and which can share intermediate data via the \texttt{INTERPROCESS\_STREAM} interface among themselves, as shown in Fig.~\ref{fig:spf}. Each session, with its temporary data store and processes, is self-contained and securely disposes of sensitive data upon completion at very least. Depending on privacy needs, sessions may operate in secure environments such as confidential VMs, enclaves, or containers.
Upon receiving \texttt{TX\_DATA} and \texttt{RX\_DATA}, SPF sessions aggregate data for a unified view, then process it for specific requests and use cases, such as adding semantic information about the environment. The UC data storage holds temporary data and interim results. The disclosure component ensures privacy controls are in place before releasing results to the app via the NEF in \texttt{RESULT\_STREAM}, saving to \texttt{RECORD\_STREAM}, or transferring to other sessions.

\section{Threat Analysis} \label{sec:threat_analysis}

\subsection{Trust Domains}
\begin{figure}
    \centering
    \includegraphics[width=\textwidth]{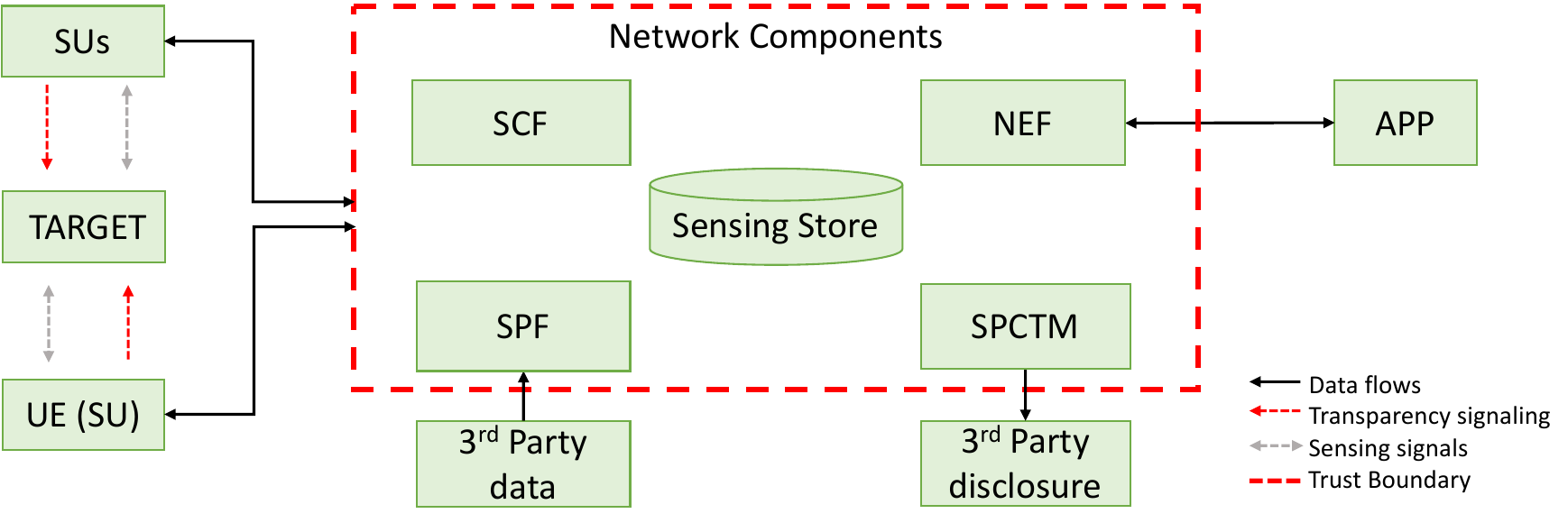}
    \caption{JCAS components and interfaces considered for threat analysis.}
    \label{fig:JCAS_threat_arch}
\end{figure}

Based on the ownership, roles, responsibilities, and access requirements, different trust boundaries are defined in the considered JCAS architecture, as shown in Fig.~\ref{fig:JCAS_threat_arch}. The trust boundaries separate the architecture into following trust domains: application, third party, target, SU, UE(SU), and network components. Moreover, as the network components might be administered by different entities, we also examine the interfaces between the sensing functions in our threat analysis. 

\subsection{Threat Models}

Both STRIDE (Spoofing, Tampering, Repudiation, Information disclosure, Denial of service (DoS), Elevation of privilege) \cite{STRIDE} and LINDDUN (Linkability, Identifiability, Non-repudiation, Detectability, Data disclosure, Unawareness and unintervenability, Non-compliance) \cite{LINDDUN} models have been utilised for security and privacy threat analysis, leveraging their widespread adoption and systematic threat identification in conceptual architecture modelling. While STRIDE focuses on security threats compromising confidentiality, integrity, or availability, LINDDUN examines potential privacy implications associated with personally identifiable information during data processing activities. The selected threats for analysis include spoofing, tampering, repudiation, information disclosure, denial of service, elevation of privilege, linkability, identifiability, detectability, unawareness, and non-compliance.

\subsection{Threats}

Appendix~\ref{sec:Appendix_threat_table} provides a summary of the threats in the proposed JCAS architecture, considering the interfaces discussed in Section~\ref{sec: prop_jcas_arch}. 
We specify the standard common weakness enumeration (CWE) \cite{CWE} in the threat table to enhance clarity and precision by providing standardised identification of weaknesses associated with each threat. 
The subsequent discussions give detailed analysis of the threats associated with each interface. Initially, our analysis focuses on external interfaces to assess potential threats. Subsequently, we evaluate threats from intra-network interfaces, considering distinct trust boundaries for each network function.

\subsubsection{Threats in Application  $\xleftrightarrow{} $ NEF Interface}


Due to the presence of sensing responses, the interface between NEF and applications poses a potential target for diverse threats. \textit{Linkability} threats may emerge if attackers can correlate multiple \texttt{SENS\_REQ} and \texttt{SENS\_RESPONSE} exchanges, potentially exposing sensitive behaviours or patterns. Insufficient security measures for protecting authentication tokens and session identifiers exchanged between the NEF and the application pose risks of \textit{identifiability} and \textit{spoofing}. 
Moreover, interception of PIIs of the NEF could allow attackers to impersonate the NEF and send malevolent responses to applications. Additionally, improper data protection mechanisms may expose sensitive sensing data in the \texttt{RESULT\_STREAM} to eavesdropping.

Attackers may conduct \textit{DoS} attacks by overwhelming the NEF with numerous \texttt{SENS\_REQ}, disrupting services for legitimate applications. Similarly, a high volume of seemingly genuine \texttt{SENS\_RESP} and \texttt{RESULT\_STREAM} could impact applications. By manipulating sensing requests and responses, attackers can tamper with control flows, inject false commands, or alter communication protocols, potentially leading to unauthorised actions or service degradation. Tampered \texttt{RESULT\_STREAM} could cause inaccurate decision-making or operational disruption. Additionally, tampering with digital signatures in \texttt{SENS\_REQ}, \texttt{SENS\_RESP}, and \texttt{RESULT\_STREAM} messages undermines \textit{non-repudiation}, allowing attackers to repudiate legitimate transactions if private keys are compromised.


If access control policies are inadequately designed, an application with an \textit{elevation of privilege} can request sensing data from unauthorised environments or restricted areas like military zones. Failure to communicate policy changes, such as data categorisation or PII classification, to the application and NEF, may inadvertently disclose sensitive information in sensing requests or responses. \textit{Non-compliance} and policy violations may grant attackers access to sensitive data from application sensing sessions, potentially involving legitimate NEF and application participation in illegitimate sensing activities.

\subsubsection{Threats in SU / UE(SU) $\xleftrightarrow{} $ Network Interface}

The wireless nature of this interface renders it susceptible to numerous threats. Attackers could \textit{link} the sensing environment, units involved, and requesting applications from exchanged radio configuration and sensing information. Inadequate confidentiality measures may \textit{disclose} sensitive data like sensed targets, areas, involved units, and network ingestion points. Furthermore, session identifiers used for radio session configuration may provide \textit{identifiable} information about entities participating in specific sessions.

By \textit{spoofing} the PIIs of the SCF, an attacker may send unauthorised radio configuration requests \texttt{CONFIG\_RADIO\_SESSION} to SUs and collect sensing data. Alternatively, if PIIs of the SU are disclosed, the attacker can send malicious data and acknowledgements to the network, potentially ensuring non-repudiation to appear legitimate. \textit{Elevated privileges} could allow the attacker or SCF to control SU sensing sessions and direct them for unlawful activities. \textit{Tampering} with sensing data and injecting malicious programs can disrupt final results and network components. Attackers may conduct \textit{DoS} attacks by flooding the sensing unit with seemingly legitimate \texttt{CONFIG\_RADIO\_SESSION} messages or jamming to block them, preventing radio sensing initiation. If the SU is integrated into the gNB, it can influence the communication system, including signalling processes.

\textbf{When SU resides in UE: } If the SU is part of the UE, other forms of information besides captured signals from targets are exchanged in the interface, assuming that the UE can process sensing information. Metadata can \textit{disclose} details about sensing sessions, allowing adversaries to link sessions from the application with metadata transferred from the UE. By \textit{elevating privileges}, the SCF may enable sensitive features like GPS and IP localisation of the UE, potentially leading to unauthorised tracking. With \textit{elevated privileges}, the UE may unknowingly engage in sensing activities, necessitating the imposition of sensing policies and regulations. Additionally, standards for handling PIIs should be followed and properly communicated to the UE. Due to the UE's involvement in various processing tasks, the SCF can obtain metadata about the UE's storage and processing capabilities.

\subsubsection{Threats in Target $\xleftrightarrow{} $ SU Interface}
The sensing units emit signals and capture reflections from targets, allowing adversaries to establish connections and potentially \textit{disclose} sensitive information about target movements or activities. Additionally, adversaries might uniquely \textit{identify} or \textit{link} individual targets from observations, jeopardising target anonymity and operational security. Attackers could \textit{impersonate} legitimate targets or deceive sensing units with false signals, leading to misidentification or inaccurate target tracking. Threats during sensing operations may include \textit{DoS} attacks aimed at disrupting radar functionality, lack of awareness or \textit{non-compliance} with access control measures, and violations of privacy regulations, compromising target security and privacy. Similarly, failure to comply with privacy regulations or ethical guidelines may lead to unauthorised collection, storage, or sharing of personal or sensitive information, violating privacy rights.

The integration of sensing units into UEs amplifies privacy concerns, as UEs with sensing capabilities may inadvertently gather sensitive data about individuals or their surroundings. For instance, SU-enabled UEs could potentially monitor users' movements or activities, heightening privacy risks if this data is misused or accessed without authorisation. Conversely, UEs with integrated sensing functionality could be exploited for unauthorised surveillance by malicious actors, presenting privacy threats to the UEs. Attackers could also correlate UEs and their movements with sensed areas or targets during specific sessions, as well as the associated application and network information.

\subsubsection{Threats in Network $\xleftrightarrow{} $ Third-Party Interface}
The incoming malicious or erroneous third-party data may be misleading for the final sensing outcomes of the network and with intention of tampering, the third party could disrupt the network services. Adversaries might launch \textit{DoS} attacks targeting the network interfaces to the third parties, flooding with excessive data or causing network congestion. On the other hand, the network can also do such type of attack by sending malicious information to the third parties.

Security weaknesses in interfaces or integration points between the network and third-party data sources may be exploited by attackers to gain unauthorised access to sensitive data, manipulate sensor readings, and inject malicious content. From the interface, the adversary could get some \textit{identifiable} information and \textit{link} the third parties involved with the network, type of sensing data shared by a third party, and some PIIs for sensing processing session. Failure to comply with data protection regulations, privacy laws, or industry standards governing the handling and sharing of third-party data may result in inadequate security measures, leading to data breaches or unauthorised access to sensitive third-party data.
 
\subsubsection{Threats in NEF $\xleftrightarrow{} $ SCF Interface}

In this interface, a malicious entity can \textit{pretend} to be NEF and send sensing request to SCF if insecure and improper authentication policies are used. Additionally, unauthorised modification can be done to sensing request and response if sufficient mechanisms are not implemented to protect the integrity of communication. Further, an adversary can \textit{observe} the communication and can find out sensitive information about sensing request, such as which entity (APP) is requesting sensing, what is the granularity, and target area of sensing etc. 


The adversary can also observe the attributes of $REQ$ and $RESP$ messages and potentially \textit{link} or, in the worst case, \textit{identify} the entities requesting sensing, sensing targets, and network components involved in sensing. \textit{DoS} threats are also applicable to NEF and SCF communication, as an adversary can \textit{tamper} with the \texttt{JCAS\_API\_REQ} so that SCF is unable to process the sensing request. This can be achieved either through unsupported fields and parameters or by creating a very large sensing request. Volumetric DoS attacks are also possible if the adversary attempts to flood SCF and NEF with a massive number of \texttt{JCAS\_API\_REQ} and \texttt{JCAS\_API\_RESP} messages, respectively. It should be noted that improper access control on either side can result in an entity being able to access resources that it is not authorised to use. 
It is important for NEF to work in \textit{compliance} with standards and policies and not initiate any unauthorised sensing requests. For example, malicious insider or operator may conduct non-compliant sensing for financial or political motives. 

\subsubsection{Threats in SCF $\xleftrightarrow{} $ SPF Interface}

An adversary can \textit{spoof} SCF and request SPF to process a sensing request, initiating unauthorised actions such as scan area requests or object tracking. Without secure communication between SCF-SPF, message \textit{tampering} is possible, allowing the adversary to alter critical parameters like ingress points (port address where SU should send the sensing data) and sensing processing information (e.g., size, duration, frequency) etc. \textit{DoS} threats on the SCF-SPF communication channel can occur through crafting large \texttt{PROC\_REQ} or flooding with numerous requests of this type, filling up SPF's capability to serve these requests and exhausting its resources. \textit{Disclosed} information from insecure messages, like network ingestion points or sensing processing characteristics, can aid subsequent attacks.

An entity at SCF can \textit{elevate its privileges} to access SPF resources, including SPF sessions, SPF ephemeral data store, and processing pipeline, allowing it to arbitrarily start and end a sensing processing session if proper access control mechanisms are not implemented. By analysing the attributes of messages exchanged between SCF and SPF, as well as other collected messages, an adversary can \textit{link} relevant data flows and \textit{identify} sensitive information about targets, UE identities, app identity, geo-location, etc. \textit{Repudiation} threats may also arise if communication logging is insufficient. At a lower level, an adversary can perform traffic fingerprinting by observing entity traffic, extracting sensing attributes in a given deployment. Similarly, a number of \textit{tampering} threats could lead to \textit{non-compliance} if policies at SCF/SPF are tampered with. 

\subsubsection{Threats in NEF $\xleftrightarrow{} $ SPF Interface}
SPF transmits the result of a sensing session (\texttt{RESULT\_STREAM}) to the requesting entity APP via NEF, which ensures the application of correct sensing policies to the result. Numerous threat categories are possible to the communication between NEF and SPF. Unauthorised entities may \textit{impersonate} SPF and transmit manipulated or malicious sensing results to APP. Similarly, unauthorised alterations to \texttt{RESULT\_STREAM} could lead to the injection or removal of objects in the sensing result. A \textit{privilege escalation} attack at NEF could bypass policy checks and expose sensing results to unauthorised entities, including NEF, internal entities, or external applications. Failure to apply required privacy controls as per the sensing policy to the \texttt{RESULT\_STREAM} may result in \textit{non-compliance} threats within the network. Moreover, an adversary monitoring data on this interface could \textit{identify} or \textit{link} sensitive information about sensing entities, target environments, and sensing activities within the environment.

\subsubsection{Threats in SPCTM $\xleftrightarrow{} $ SCF/SPF/NEF Interface}


Communication between SPCTM and other functions involve sensitive information, such as sensing policies, consent information, and disclosure logs. Therefore, the interfaces that involve SPCTM, can be the target for many threats. Without proper authentication mechanisms and encryption, unauthorised entities may \textit{spoof} \texttt{SENS\_POL\_REQ} to SPCTM, masquerading as legitimate NEF, to ascertain what an APP/entity is authorised to sense. Additionally, the interface may \textit{disclose} sensitive information, such as sensing policies, the type of sensing requested by an entity/APP, and comprehensive logs of sensing requests and results. Moreover, adversaries could extract sensitive information, such as processing policies corresponding to a subject/APP/area, results of sensing sessions, or valuable data (e.g., IP addresses of operators), leading to \textit{linkability} and \textit{identifiability} threats. \textit{Non-compliance} threats may arise from incorrect functioning or malicious tampering at SPCTM, potentially resulting in the retrieval of incorrect sensing and processing policies in SPF, SCF, and NEF.

\section{Discussion and Mitigation Strategies} \label{sec:discussion}

\subsection{Privacy Enforcement through SPCTM and SPF}

In this work, we proposed SPCTM as the governing entity for the privacy framework in JCAS, supporting mechanisms for transparency, consent management, sensing policy management, and responsible accountability, along with authentication and authorisation. Although SPCTM does not directly enforce privacy controls itself, it serves as a governance framework that oversees various aspects of privacy and regulatory requirements. Transparency data, coupled with the sensing transparency function, informs users about their involvement in sensing activities, complying with GDPR requirements for the right to be informed and ensuring awareness of ongoing sensing activities. Similarly, consent data managed by sensing consent management function ensures that sensing activities respect the consent of affected data subjects, while also adhering to the latest consent policies during sensing and data processing. Sensing policies and current session policies stored at SPCTM collaborate with the sensing policies decision point function, enabling all sensing functions (SCF and SPF) to comply with the most recent sensing and session policies. Additionally, NEF utilises the SPD point to authorise an application or entity for a given sensing request, while also maintaining sensing-related logs at SPCTM for accountability and transparency purposes.  


On the other hand, we proposed short-lived sensing processing session in SPF. These ephemeral sessions in SPF are orchestrated by SPF with privacy controls, which are applicable on disclosure as per the sensing policy. 
These privacy controls ensures that the data being written to the sensing store safeguards the PIIs. Similarly, before sending the sensing result (\texttt{RESULT\_STREAM}) to app (through NEF), necessary privacy controls should be applied to the result to disclose only necessary information to app in a secure way. While not solving all privacy concerns, properly configured transient SPF sessions offer a mechanism to address numerous privacy issues. 

\subsection{Suggested Security and Privacy Controls}


Here, we suggest some of the essential security and privacy controls for JCAS system. To minimise the overall negative effect on system's functionality, dependability, safety, and to improve cost efficiency, we advocate for the security and privacy solutions that are the most necessary for mitigating the threats in \cref{sec:Appendix_threat_table}.

\subsubsection{Identification and Authentication}

Establishing high levels of assurance for the information originating and communicated by the components, modules, and interfaces in JCAS necessitates ensuring the authenticity of that information. The latter usually use corresponding hardware and software roots of trust inherent to these components, modules, and interfaces \cite{YANG2020629,root_tr01}. Further steps in establishing authenticity can involve various tools and methods.

Identification and authentication in JCAS, whether hardware or software modules, must establish and verify claimed identities and associated security attributes. Due to the involvement of sensitive entities, such as targets and the SUs and their sensitive information, solutions should validate entity identities, their authority to interact with other entities, and ensure correct association of security attributes \cite{4090091}. Furthermore, solutions should establish parameters such as the maximum number of unsuccessful authentication attempts to reduce the vulnerability to brute-force attacks. It is also suggested that suitable actions for failures, such as entity lockout or triggering alerts for further investigation, be outlined to enhance system security.
Attention should be paid to authentication mechanisms supported by JCAS components and modules and the properties of the attributes on which they are based. For example, unforgeable authentication prevents the forging or copying of authentication data, while single-use mechanisms operate with data for one-time use \cite{SU201411}. 


\subsubsection{Data Protection}

Ensuring the integrity, confidentiality, and access control of signals, sensed data, and associated security attributes within JCAS modules is paramount. This entails establishing policies for data protection, implementing techniques for data flow protection, and managing offline storage, import, and export procedures.


\begin{itemize}

\item \textbf{Access and information flow control}: Access control policies define specific security behaviours enforceable within JCAS, outlining requirements for relevant security techniques and means. This control scope involves three main elements: the subjects and objects governed by the policy and the operations covered by it \cite{10.1145/504909.504910,10.1145/1242572.1242664}. Information flow control policies delineate control scope, characterised by three sets: controlled subjects, controlled information, and operations governing information flow to and from controlled subjects \cite{10.1145/1554339.1554352}. For instance, as proposed in this paper, the SPCTM component (see \cref{fig:JCAS_threat_arch}) incorporates access control policies and information flow control policies.
\item \textbf{Information retention and disposal}: We recommend implementing information retention control in JCAS to securely manage data that is no longer needed by components or modules. This includes deleting copies of specified objects or data when they are no longer necessary for operation and defining necessary operations for each object \cite{6335436}. Additionally, residual information protection ensures that data in a resource is not accessible when the resource is de-allocated from one entity and reallocated to another, preventing data leakage. Furthermore, it is necessary to safeguard data stored in a resource that has been logically deleted or released but may persist within the controlled resource, potentially being reallocated to another object \cite{JORDON201212}.


\item \textbf{Integrity and confidentiality of sensing data-in-store}: To protect the integrity of sensing data and associated PIIs, we suggest implementing rollback solutions, which revert the last operation or a sequence of operations within a defined limit, such as a time period, and restore to a previously known state \cite{5658636,r_back_01}. Integrity solutions for data stored in the sensing store should safeguard it within a component or module, monitoring and correcting errors that may impact data stored in memory or storage devices \cite{10.1145/1103780.1103784}. We suggest the confidentiality of sensing data stored in the JCAS components and modules to restrict access to memory data solely through specified interfaces and prevent unauthorised information access. The specifics of these confidentiality solutions may vary based on designated memory areas, cryptographic methods, or the necessity of JCAS stakeholder intervention \cite{10.1145/2566673}.  


\item \textbf{Integrity and confidentiality of signals-and-data-in-transit}: To ensure the integrity of control signals and sensed data transmitted across JCAS interfaces, we propose employing solutions capable of detecting modifications, deletions, insertions, and replay errors \cite{wicker1995error}. Additionally, for data recovery at the receiving end, options include utilising source assistance (e.g., automatic repeat query) or standalone recovery methods (e.g., forward error correction) \cite{9471818}. Moreover, it is imperative to implement confidentiality-based solutions to safeguard JCAS data from disclosure during transit.


\item \textbf{Security of imported/exported sensing data}: We recommend employing solutions for data authentication, export, and import to secure offline operations on sensing data. Data authentication allows an entity to verify the authenticity of the information, ensuring the validity of specific data units and preventing forgery or fraudulent modification \cite{10.1007/978-3-319-94478-4_14}. Depending on the use case, exporting data from the JCAS component or module (e.g., to a third party) should either maintain security attributes and data protection or discard them post-export. This security feature focuses on export limitations and the association of security attributes with exported user data \cite{d10.1145/1314466.1314477}. Similarly, security techniques for importing data into JCAS (e.g., from third parties) must address import limitations, define desired security attributes, and interpret associated security attributes with the imported data \cite{8272322}.

\end{itemize}

\subsubsection{Privacy Controls}

Here, we discuss the core privacy controls and extended privacy controls. Core privacy controls safeguard an entity's PII from discovery and misuse, encompassing anonymity, pseudonymity, unlinkability, and unobservability. We also emphasise the need for extended privacy controls, including consent-and-transparency-enabling policies concerning sensing activities. The following details of the core privacy controls shall be considered. 



\begin{itemize}
    \item \textbf{Anonymisation and pseudonymisation}: During sensing, techniques such as data aggregation, randomisation, and masking, could obscure identifying information of targets and sensing units during sensing data processing\cite{10.1145/1103576.1103585,10.1007/978-3-319-46963-8_1}. Different encryption and tokenisation approaches can further safeguard sensitive data while transmitting sensing signals from the SU to the SPF. Further, differential privacy methods add noise to data, preserving statistical properties, while dynamic pseudonymisation assigns temporary identifiers to prevent long-term tracking. Implementing these measures can ensure privacy in sensing systems throughout data processing, transmission, and analysis, allowing for valuable insights while protecting individual privacy \cite{survey_privacy}.

    \item \textbf{Unlinkability and unobservability}: Unlinkability of sensing requests from the applications and responses is essential for upholding privacy rights, for instance, preventing unauthorised tracking \cite{10.1007/978-3-540-40956-4_3}. Further, configuration messages from the SCF to the SUs and the responses from the SUs need methods to ensure unlinkability. When applications make sensing requests, unlinkability may be achieved by anonymising the requester's identity through techniques like dynamic pseudonymisation or session IDs. During sensing activities, unlinkability may be achieved by obfuscation methods to introduce randomness into data, data fragmentation for making the reconstruction process challenging, and employing cryptographic or physical layer security methods to protect data transmission. This ensures that data collected from targets cannot be easily linked back to specific individuals or entities. 
    Unobservability of sensing units and network resources involved in sensing may be achieved through covert operations and channels, noise injection, anonymisation, data fragmentation, decoy traffic, mix networks \cite{Mixnet}, and differential privacy. These methods, whether used individually or in combination, aim to conceal and confuse sensitive sensing activities or data, thereby safeguarding privacy.

    \item \textbf{Extended privacy controls}: Privacy solutions for ``consent and choice'' ensure appropriate handling of PIIs in JCAS, specifying methods, timing, and conditions for processing within corresponding modules and components \cite{10.1145/1837110.1837126}. For instance, consent initiation may occur within the SPCTM component as depicted in Fig.~\ref{fig:JCAS_arch}. To comply with the principle of ``openness, transparency, and notice'' in sensing, stakeholders should have access to general information regarding the handling of PII policy. JCAS components and modules must implement appropriate solutions to inform relevant stakeholders about any changes to the policy \cite{EVA2022107334,8510982}. In our proposal, SPCTM is considered for the above functions.
\end{itemize}

\subsubsection{Security and Privacy Controls for UE}
\label{secuesu}
\begin{figure}
    \centering
    \includegraphics[width=0.8\linewidth]{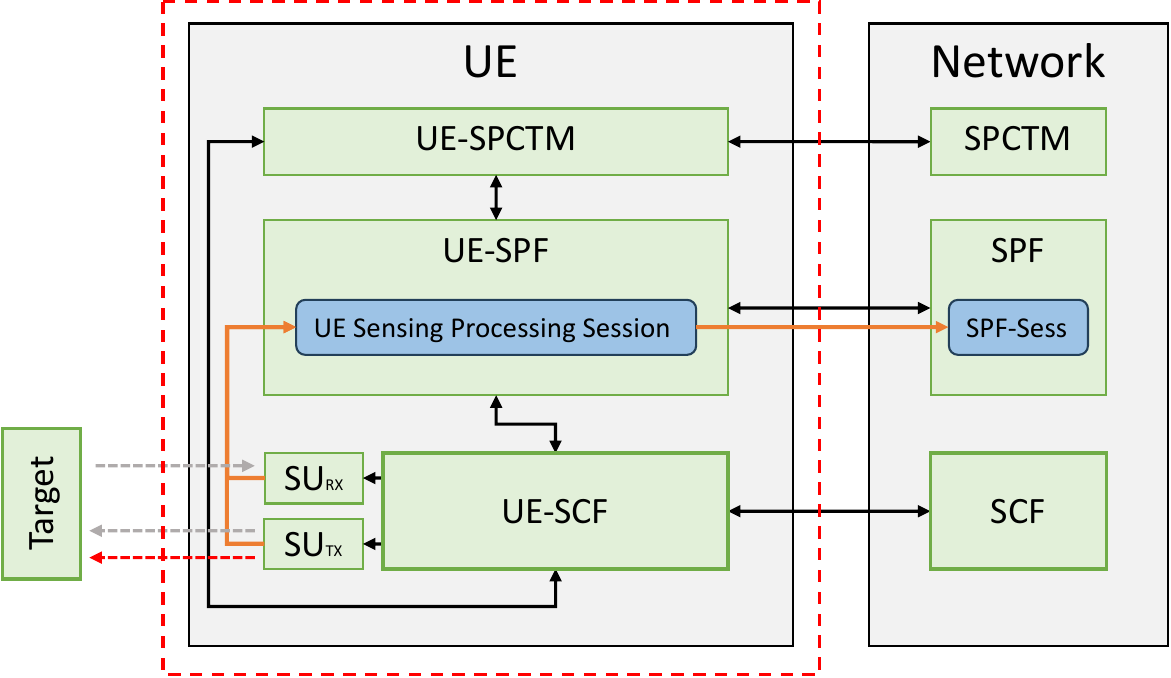}
    \caption{Illustration of UE specific controls and processing for sensing.}
    \label{fig:su-control}
\end{figure}

When user equipment integrates sensing units, it is crucial to enforce security and privacy controls on the UE to safeguard sensitive sensing data and preserve the UE's interests in consent and policy management. Functionalities similar to SPCTM on the UE side are necessary to manage sensing policies, obtain consent to enable UE sensing services and ensure transparent data disclosure to the network. Additionally, due to the involvement of UE resources in processing sensing data, network functions similar to SCF and SPF on the network side are necessary for sensing management in UE. Fig.~\ref{fig:su-control} illustrates our envisioned interfaces and functions on the UE side concerning JCAS.

Further, the controls on UE have to include data encryption during transmission and storage, strict access control mechanisms for enabling sensing services in the UE, and authentication requirements. Anonymisation and pseudonymisation techniques have to safeguard the PII and sensing information. With proper consent management, UEs can gain control over the sensing units, and logging will allow maintaining records of the sensing activities \cite{ISO_secpriv}. Implementing secure communication protocols and conducting regular privacy impact assessments can further enhance security and privacy measures, ensuring compliance with relevant regulations and standards.

\section{Conclusions and Future Work} \label{sec:conclusion}
In this paper, we examined the architectural, security, and privacy aspects of JCAS, a vital technology anticipated in 6G networks. Drawing from our analysis on the emergent JCAS architecture, we proposed some enhancements involving additional network functions and interfaces to address key challenges related SPCTM and privacy of PIIs involved in sensing. Subsequently, we performed a detailed threat analysis for each interface within the proposed JCAS architecture, aligning with standard CWEs and presenting a threat summary table synchronised with JCAS threats. To mitigate the security and privacy risks associated with JCAS, we put forth security and privacy controls, emphasising their significance for JCAS systems.

As part of the future extension of this work, we intend to conduct a more comprehensive risk assessment of the proposed JCAS architecture, including a detailed analysis of threat likelihood and potential impact. We will explore how threats to sensing activities affect the communication systems. Additionally, we plan to propose fine-grained privacy controls within the JCAS framework to ensure stronger protection of user data and privacy rights. Furthermore, we intend to develop detailed mitigation techniques to effectively address identified risks, providing actionable strategies to enhance the security and privacy of JCAS systems.

\section*{Acknowledgement}

This work has been partly funded by the European Commission through the project Hexa-X-II (Grant no.
101095759). Additionally, the authors from Bark\-hau\-sen Institute are supported by the Federal Ministry of Education and Research, Germany (Grant no. 16KISK231, 16KISK122), and are also financed based on the budget passed by the Saxonian State Parliament in Germany.



\appendix
\section{Summary of Threats} \label{sec:Appendix_threat_table}

In accordance with the CWE structure, the CWEs specified in the table are at the class-level to avoid being overly generalises or excessively specific.

\begin{table} [!ht]
  \centering
   \renewcommand{\arraystretch}{1.7} 
    \caption{Summary of threats in the proposed JCAS architecture}
    \scalebox{0.8}{%
    \begin{tabular}{|m{6cm}|m{5cm}|m{4cm}|} 
        \hline
        \textbf{Weakness or reason} & \textbf{Threat type} & \textbf{Target component} \\
        \hline
        Inadequate protection mechanism for the hardware resources of the JCAS devices (CWE 1263) & 
        Linkability, Identifiability, Detectability, Data disclosure, Tampering, Spoofing  & 
        SU, UE(SU), APP, network hardware \\  
        \hline  
        Improper authentication, authorisation, ownership and privilege management (CWEs: 287, 269, 282, 285) & Linkability, Identifiability, Detectability, Data disclosure, Unawareness, Spoofing, Tampering &
         SU, UE(SU), APP, targets, NEF, SCF, SPF, SPCTM, third parties \\   
        \hline 
        Exposure of sensitive sensing and communication information to the actors from unauthorised trust domain (CWEs: 668, 669)   & 
        Linkability, Identifiability, Detectability, Data disclosure, Unawareness &
        SU, UE(SU), APP, Targets, NEF, SCF, SPF, SPCTM, third parties   \\  
        \hline 
        Insecure storage of sensitive information with improper access rights (CWEs: 922,311,326) & 
        Linkability, Identifiability, Detectability, Data disclosure, Spoofing, Tampering, Unawareness &
        UE(SU), SPCTM     \\   
        \hline 
        Missing of encryption method or use of methods with inadequate strength (CWEs: 311,326, 327, 330) & 
        Linkability, Identifiability, Data disclosure &
        SU, UE(SU), APP, Targets, NEF, SCF, SPF, SPCTM, thid parties   \\
        \hline 
        Improper isolation or compartmentalisation of resources \; \; \; (CWE 653) & 
        Linkability, Identifiability, Detectability, Data disclosure, Unawareness, Spoofing, Tampering &
        UE(SU), NEF, SCF, SPF, SPCTM   \\   
        \hline
        Insufficient policy description, insufficient adherence to required conventions and sensing policies, violation of security and privacy policies (CWEs: 1059, 1076, 1177, 1061, 573, 657, 684, 758, 1357) & 
        Non-compliance &
        SU, UE(SU), APP, Targets, NEF, SCF, SPF, SPCTM, third parties \\   
        \hline
        Lack of mechanism to validate and verify the JCAS sensing request/control signal/acquired data (CWEs: 20, 74) 
         &
         Tampering, Elevation of Privilege &
        UE(SU), SU, SPF, NEF, APP \\   
        \hline
        Lack of authenticity in request/control signal, acquired data (CWEs: 287, 345)&
        Spoofing, Tampering, Elevation of privilege, Repudiation &
        UE(SU), SU, SPF, NEF, APP \\   
        \hline
        Insufficient availability and confidentiality of sensing request/control signal/acquired data (CWEs: 221, 404, 665, 923) &
        Data Disclosure, DoS &
        UE(SU), SU, SPF, NEF, APP \\   
        \hline
        Inadequate monitoring and management of hardware resources and data (CWEs: 400, 404, 668, 923, 1317) &
        Spoofing, Tampering, Elevation of privilege, Data disclosure, DoS &
        UE(SU), SU, SPCTM, SPF, SCF, NEF, APP \\   
        \hline
        Improper protection for time synchronisation mechanism in SU (CWE 662) & Tampering, DoS & SU, UE(SU), SPF  \\   
        \hline
        Improper logging, insufficient privacy control on disclosure logs (CWEs: 668, 778) &
        Linkability, Identifiability, Detectability, Repudiation, Data disclosure &
        SPCTM, NEF, APP, third parties \\   
        \hline
        Lack of physical layer security mechanism for jamming/physical layer attacks &
        DoS &
        UE(SU), Targets, APP, NEF, SCF, SPF \\
        \hline
        Insufficient mechanism to inform user/targets about the sensing, usage of sensing data (CWEs: 285, 668) &
        Unawareness &
        UE(SU), Targets \\
        \hline
        Incapable algorithms for load balancing, traffic filtering, and network resilience (CWEs: 400, 665) &
        DoS &
        SPF, UE(SU), NEF, APP, Third parties \\
        \hline
    \end{tabular}}
\end{table}
\clearpage 

\end{document}